\documentclass[prl,aps,twocolumn,floatfix,superscriptaddress]{revtex4-1}

\usepackage{mathrsfs}
\usepackage{amsfonts}
\usepackage{amsmath,amssymb,bm}
\usepackage{array}
\usepackage{verbatim}
\usepackage{epsfig}
\usepackage{graphicx}
\usepackage[colorlinks=true,linkcolor=blue]{hyperref}
\usepackage[normalem]{ulem}
\usepackage{xcolor}
\usepackage{ulem}

\def\Phperp{P_{h\perp}}
\def\kperp{k_\perp}
\def\pperp{p_\perp}
\def\bfPhperp{{\bm P}_{h\perp}}

\def\bfpperp{{\bm p}_\perp}

\def\avkperp{\la \kperp^2 \ra}
\def\avpperp{\la \pperp^2 \ra}
\def\bfhp{\hat{\bm h}}
\newcommand{\la}{\langle}
\newcommand{\ra}{\rangle}

\newcommand{\fref}[1]{Fig.~\ref{f.#1}}

\begin{document}

\preprint{MSUHEP-17-016,JLAB-THY-17-2574}

\title{First Monte Carlo global analysis of nucleon transversity
        with lattice QCD constraints}

\author{H.-W. Lin}      % \email{hwlin@pa.msu.edu}
\affiliation{Michigan State University, East Lansing, Michigan 48824, USA}

\author{W. Melnitchouk} % \email{wmelnitc@jlab.org}
\affiliation{Jefferson Lab, Newport News, Virginia 23606, USA}

\author{A. Prokudin}    % \email{prokudin@jlab.org}
\affiliation{Jefferson Lab, Newport News, Virginia 23606, USA}
\affiliation{Penn State Berks, Reading, Pennsylvania 19610, USA}

\author{N. Sato}	% \email{nsato@jlab.org}
\affiliation{University of Connecticut, Storrs, Connecticut 06269, USA}

\author{H. Shows~III}   % \email{}
\affiliation{Louisiana State University, Baton Rouge, Louisiana 70803, USA\\
	\vspace*{0.2cm}
	{\bf Jefferson Lab Angular Momentum (JAM) Collaboration
	\vspace*{0.2cm} }}

\begin{abstract}
We report on the first global QCD analysis of the quark transversity
distributions in the nucleon from semi-inclusive deep-inelastic
scattering (SIDIS), using a new Monte Carlo method based on nested
sampling and constraints on the isovector tensor charge $g_T$ from
lattice QCD.
A simultaneous fit to the available SIDIS Collins asymmetry data is
compatible with $g_T$ values extracted from a comprehensive reanalysis
of existing lattice simulations, in contrast to previous analyses which
found significantly smaller $g_T$ values.
The contributions to the nucleon tensor charge from $u$ and $d$
quarks are found to be $\delta u =  0.3(2)$ and $\delta d = -0.7(2)$
at a scale $Q^2 = 2$~GeV$^2$.
\end{abstract}

\date{\today}
\maketitle

%%%%%%%%%%%%%%%%%%%%%%%%%%%%%%%%%%%%%%%%%%%%%%%%%%%%%%%%%%%%%%%%%%%%%%%%
% Introduction.
%
Along with the unpolarized ($f_1$) and helicity-dependent ($g_1$)
parton distribution functions (PDFs), the transversity distribution
($h_1$) completes the full set of quark PDFs that characterize the
collinear structure of the nucleon at leading twist.
While considerable information has been accumulated on the first two
distributions from several decades of deep-inelastic scattering (DIS)
and other high-energy experiments~\cite{Gao:2017yyd,
Blumlein:2012bf, Jimenez-Delgado:2013sma, Aidala:2012mv},
comparatively little is known about the transversity PDFs.
The transversity PDF, $h_1^q(x)$, gives the distribution of a
transversely polarized quark $q$ carrying a momentum fraction
$x$ in a transversely polarized nucleon, and its lowest moment,
	$\delta q \equiv
	 \int_0^1 dx [h_1^q(x) - h_1^{\bar q}(x)]$,
gives the nucleon's tensor charge for quark $q$~\cite{Ralston:1979ys,
Jaffe:1991kp, Cortes:1991ja, Barone:2001sp, Accardi:2012qut,
Ye:2016prn, Accardi:2017pmi}.
In addition to providing fundamental information on the quark spin
structure of the nucleon, the tensor charge also plays an important
role in constraining hadronic physics backgrounds in probes of physics
beyond the Standard Model~\cite{DelNobile:2013sia, Bhattacharya:2011qm,
Courtoy:2015haa}.

Compared with the chiral-even $f_1$ and $g_1$ PDFs, the experimental
exploration of the chiral-odd $h_1$ is considerably more involved,
requiring the coupling of the transversity distribution to another
chiral-odd function~\cite{Jaffe:1991kp}.
Observables sensitive to transversity include the Collins single-spin
asymmetries in semi-inclusive deep-inelastic scattering (SIDIS), where
$h_1$ couples to the chiral-odd Collins fragmentation function (FF)
$H_1^\perp$~\cite{Collins:1992kk}, while two Collins FFs generate an
azimuthal asymmetry in two-hadron production in $e^+ e^-$
annihilation~\cite{Boer:1997mf}.

Several previous analyses have attempted to extract the transverse
momentum dependent (TMD) transversity distributions, from both
SIDIS and $e^+ e^-$ data.
Anselmino {\it et~al.}~\cite{Anselmino:2007fs, Anselmino:2008jk,
Anselmino:2013vqa} employed a factorized Gaussian ansatz to relate
the TMD distributions to the $h_1^q$ PDFs, while
Kang~{\it et~al.}~\cite{Kang:2014zza, Kang:2015msa} used in
addition the TMD evolution formalism~\cite{Collins:2011zzd}.
In both cases the $x$ dependence of $h_1^q(x)$ was parametrized
in terms of the sum of unpolarized and helicity distributions
at the initial scale.
Working within collinear factorization, Bacchetta {\it et~al.}
\cite{Bacchetta:2012ty, Bacchetta:2011ip, Radici:2015mwa} also
extracted transversity PDFs from pion pair production in SIDIS
using dihadron FFs from $e^+ e^-$ data.
These analyses gave values for the isovector moment
	$g_T \equiv \delta u - \delta d$
in the range $0.4 - 1$, with sizeable ($30\%$--50\%) uncertainties.
In all of these studies, the experimental coverage was restricted to
the region $0.02 \lesssim x \lesssim 0.3$, so that the determination
of the full moment required extrapolation outside the measured range.

Complementing the challenging empirical extractions of transversity,
first-principles lattice QCD calculations can provide additional
information on the nucleon transverse spin structure.
While recent breakthroughs in quasi-PDFs have allowed the first
direct lattice computations of the $x$ dependence of
transversity~\cite{Ji:2013dva, Chen:2016utp}, calculations of
moments of the isovector $h_1^q$ PDF are more developed,
with a number of simulations of $g_T$ having been
performed~\cite{Bhattacharya:2016zcn, Bhattacharya:2015wna,
Green:2012ej, Aoki:2010xg, Abdel-Rehim:2015owa, Bali:2014nma,
Yamazaki:2008py} at physical pion masses and with multiple
lattice spacings and volumes.
No significant contamination from excited states has been observed,
along with very mild volume and lattice spacing dependence, making
$g_T$ a ``golden'' channel in lattice nucleon structure studies.
Curiously, however, all of the simulations give values of $g_T$
close to unity, in contrast to the phenomenological values,
which are generally smaller~\cite{Kang:2015msa, Ye:2016prn},
with central values $\sim 0.5-0.6$.
This prompts the question whether the systematic differences
between the lattice and phenomenological results suggest a real
tension between the two.
From the uncertainties found by Kang {\it et~al.}~\cite{Kang:2015msa},
for example, one would conclude that, after the inclusion of data
from the future SoLID experiment at Jefferson Lab~\cite{Ye:2016prn},
the phenomenological values of $g_T$ would be incompatible with
lattice at more than $5\sigma$~C.L.

In this Letter, we address the question of whether the experimental
data on transversity are compatible with the lattice $g_T$ results
--- whether there indeed is a ``transverse-spin puzzle'', as suggested
by some of the previous analyses~\cite{Kang:2015msa, Ye:2016prn}
--- by using the lattice data on $g_T$ as an additional constraint
on the global QCD analysis of transversity.
We implement several important improvements over previous analyses,
making use of a more robust fitting methodology based on Monte Carlo
(MC) sampling methods.  Specifically, we use the nested sampling
algorithm~\cite{Skilling:2004, Mukherjee:2005wg, Shaw:2007jj},
which maps the likelihood function into an MC-weighted parameter
sample and allows a rigorous determination of PDF uncertainties.
This approach improves the fitting methodology of
Refs.~\cite{Kang:2014zza, Kang:2015msa} by allowing more flexible
parametrizations of the initial conditions of the transversity
and Collins FFs.
Similar MC-based methods have recently been used to analyze
collinear PDFs~\cite{Sato:2016tuz, Ethier:2017zbq} and
FFs~\cite{Sato:2016wqj, Ethier:2017zbq}, but have never before
been applied to TMDs.

%%%%%%%%%%%%%%%%%%%%%%%%%%%%%%%%%%%%%%%%%%%%%%%%%%%%%%%%%%%%%%%%%%%%%%%%
% Lattice QCD.
%
To begin with, we revisit the existing lattice QCD simulations of
$g_T$ to obtain a reliable averaged data point that can be used
in the global QCD analysis.
One challenge is that the various lattice calculations
estimate systematic uncertainties differently, making it problematic
to simply average the reported values.
We instead combine the available dynamical simulation data,
using only calculations with multiple lattice spacings, volumes
and quark masses; we use several procedures to ensure that the
final uncertainties are not underestimated.

There are 3 available data sets that meet these criteria:
  the PNDME Collaboration results with \mbox{$N_f=2+1+1$}
  flavors~\cite{Bhattacharya:2016zcn}
  the RQCD Collaboration data with $N_f=2$~\cite{Bali:2014nma};
  and the LHPC set with $N_f=2+1$~\cite{Green:2012ej}.
Cuts on the data are imposed for pion masses
	$m_\pi^2 < 0.12$~GeV$^2$
and for
	$m_\pi L > 3$,
where $L^3$ is the lattice volume, to control the chiral
and infinite-volume extrapolations.
Since all of the lattice simulations show a mild dependence on
the volume and lattice spacing $a$, the simplest approach is
to extrapolate $g_T$ considering only the $m_\pi$ dependence.
Extrapolating the data either linearly in $m_\pi^2$ or including
chiral logarithms ($\sim m_\pi^2 \ln m_\pi^2$), as predicted from
chiral effective theory~\cite{Chen:2001eg, Detmold:2002nf},
gives $g_T^\text{latt} = 1.006(22)$.

To further include uncertainties from taking the continuum limit,
we assign a different lattice discretization extrapolation
coefficient for each simulation~\cite{Bhattacharya:2016zcn,
Bali:2014nma, Green:2012ej}.
To account for the different actions, we use ${\cal O}(a)$
for the PNDME and LHPC results, and ${\cal O}(a^2)$ for RQCD.
For the volume dependence, we consider both $e^{m_\pi L}$
and $m_\pi^2 e^{m_\pi L}$ forms.
Taking all possible combinations then gives 12 distinct fitting
formulas for the continuum extrapolation of $g_T$.
The results of these fits are combined using the Akaike information
criterion, $\text{AIC} = 2k + \chi^2$, where $k$ is the number of
free parameters in the fit and $\chi^2$ is the minimum sum of
squared fit residuals.
Each fit is weighted by the factor
	$w_i = P_i/(\sum_j P_j)$,
where $P_j = \exp[-(\text{AIC}_j - \min\text{AIC})/2]$,
which yields $g_T^\text{latt} = 1.008(56)$.

Another approach is to average the lattice data using methods
advocated by the Flavor Lattice Averaging Group
(FLAG)~\cite{Aoki:2016frl}.
However, given that most extrapolations of nucleon matrix elements
do not explicitly control finite volume and lattice spacing
systematics, such an averaging will be dominated by results
with the most optimistic systematic uncertainty estimates.
We extrapolate, therefore, each group's data using a single, universal
formula, assuming linear dependence on $m_\pi^2$, $e^{m_\pi L}$ and
$a$ (or $a^2$), and then perform a weighted analysis as in the FLAG
approach.
The result is $g_T^\text{latt} = 1.00(5)$, which is consistent with
the above estimate.  To be conservative, we take the larger
uncertainty, $g_T^\text{latt} = 1.01(6)$, as the final averaged
value to be used in the global analysis.

%%%%%%%%%%%%%%%%%%%%%%%%%%%%%%%%%%%%%%%%%%%%%%%%%%%%%%%%%%%%%%%%%%%%%%%%
% Collins asymmetry.
%
For the experimental data used in our fit, we consider the
$\sin(\phi_h+\phi_s)$ modulation of the differential SIDIS
cross section, or Collins asymmetry,
\begin{equation}
A_{UT}^{\rm \sin(\phi_h+\phi_s)}
= \frac{2 (1-y)}{1+(1-y)^2}
  \frac{F_{UT}^{\sin\left(\phi_h +\phi_s\right)}}{F_{UU}},
\end{equation}
where $\phi_h$ and $\phi_s$ are the azimuthal angles for the
transverse momentum of the produced hadron $h$ and the nucleon
spin vector with respect to the lepton plane
in the virtual photon--nucleon center of mass frame, and $y$
is the fractional energy loss of the incident lepton.
The structure functions
   $F_{UU}$ and
   $F_{UT}^{\sin\left(\phi_h+\phi_s\right)}$ are functions
of the Bjorken variable
   $x = Q^2 / 2 P\cdot q$,
the hadron momentum fraction
   $z = P \cdot P_h / P \cdot q$, and
the hadron transverse momentum $P_{h\perp}$,
where $P$, $P_h$ and $q$ are the four-momenta of the target,
produced hadron, and exchanged photon, respectively,
and $Q^2 = -q^2$.
For $\Phperp \ll Q$ these can be written
as convolutions of the unpolarized $f_1^q$ TMD PDF and unpolarized
$D_1^{h/q}$ TMD FF, and the TMD transversity PDF $h_1^q$ and
$H_1^{\perp\, h/q}$ (Collins) FF, respectively,
\begin{eqnarray}
F_{UU}
&=& {\cal C}\biggl[ f_1 \otimes D_1 \biggr],
\label{eq:FUU}					\\
F_{UT}^{\sin\left(\phi_h + \phi_s\right)}
&=& {\cal C}\biggl[ \frac{\bfhp \cdot \bfpperp^{ }}{z m_h}
		    \otimes h_1
		    \otimes H_1^{\perp}
	    \biggr],
\label{eq:FUT}
\end{eqnarray}
where ${\cal C}$ is the standard TMD convolution operator
\cite{Bacchetta:2006tn}, $\bfhp$ is a unit vector along
$\bfPhperp$, and $\bfpperp$ the transverse momentum of $h$
with respect to the fragmenting quark.

The TMD PDFs depend on $x$ and the parton transverse momentum
$\kperp$, while the FFs depend on $z$ and $\pperp$, with their
$Q^2$ dependence governed by the Collins-Soper
equations~\cite{Collins:1981uk, Collins:2011zzd}.
The existing data on Collins asymmetries have very mild dependence
on $Q^2$ and are compatible with no evolution~\cite{Kang:2015msa,
Anselmino:2015sxa}.
For the parametrization of the unpolarized and transversity TMD
PDFs we follow Refs.~\cite{Anselmino:2007fs, Anselmino:2008jk,
Anselmino:2013vqa} in adopting a factorized form,
\begin{eqnarray}
f^q(x,\kperp^2)
&=& f^q(x)\ {\cal G}_{f}^q(\kperp^2),
\label{eq:f1h1-gauss}
\end{eqnarray}
where the generic function $f^q = f_1^q$ or $h_1^q$, and the
$\kperp^2$ dependence is given by a Gaussian distribution,
\begin{eqnarray}
{\cal G}_f^q( \kperp^2)
&=& \frac{1}{\pi\avkperp_f^q}\;
    {\exp\left[{-\frac{\kperp^2}{\avkperp_f^q}}\right]}.
\label{eq:Ggauss}
\end{eqnarray}
The transverse widths $\avkperp_{f}^q$ are in general
flavor dependent, and can be functions of $x$, although
here
we assume their $x$ dependence is negligible.
For the TMD FFs, the unpolarized distribution is parametrized
analogously,
\begin{eqnarray}
D_1^{h/q}(z,\pperp^2)
&=& D_1^{h/q}(z)\ {\cal G}_{D_1}^{h/q}(\pperp^2),
\end{eqnarray}
while the Collins FF involves an additional $z$-dependent weight factor,
\begin{eqnarray}
H_1^{\perp h/q}(z,\pperp)
&=& \frac{2 z^2 m_h^2}{\avpperp^{h/q}_{H_1^\perp}}\,
    H_{1\, h/q}^{\perp (1)}(z)\
    {\cal G}_{H_1^\perp}^{h/q}(\pperp^2).
\label{eq:coll-funct_new}
\end{eqnarray}
The $\pperp^2$ dependence of the functions
	${\cal G}_{D_1}^{h/q}$ and
	${\cal G}_{H_1^\perp}^{h/q}$
is assumed to be Gaussian, in analogy with (\ref{eq:Ggauss}),
with the average $\avpperp^{h/q}$ independent of $z$.
The $z$ dependence of the Collins FF is parametrized in terms
of its $\pperp^2$-weighted moment, $H_{1\, h/q}^{\perp (1)}(z)$
\cite{Kang:2015msa}.
Using the TMD PDFs and FFs in
Eqs.~(\ref{eq:f1h1-gauss})--(\ref{eq:coll-funct_new}),
the $\Phperp^2$ dependence in the structure functions is then
proportional to
$ \exp\left( -\Phperp^2 / \la \Phperp^2 \ra_{f,D}^{h/q} \right) $,
where
$ \la \Phperp^2 \ra_{f,D}^q = z^2 \avkperp_f^q + \avpperp_D^{h/q} $.

%%%%%%%%%%%%%%%%%%%%%%%%%%%%%%%%%%%%%%%%%%%%%%%%%%%%%%%%%%%%%%%%%%%%%%%%
% Phenomenology.
%
Our global analysis fits SIDIS $\pi^\pm$ production data from
proton and deuteron targets, including their $x$, $z$ and $\Phperp$
dependence, with a total of 106 data points from the
  HERMES~\cite{Airapetian:2010ds} and
  COMPASS~\cite{Alekseev:2008aa, Adolph:2014zba}
experiments.
This gives 4 linear combinations of transversity TMD PDFs and
Collins TMD FFs for different quark flavors, from which we extract
  the $u$, $d$ and antiquark transversity PDFs
(from 4 $x$-dependent combinations) and
  the favored and unfavored Collins FFs
(from 4 $z$-dependent combinations),
  together with their respective transverse momentum widths
(from the $\Phperp$ dependence).
We do not include lower-energy Collins asymmetry data from Jefferson Lab
on $^3$He nuclei because of concerns about the separation of the current
and target fragmentation regions at relatively low
energies~\cite{Collins:2016hqq}.

In selecting the data to be used in the fit, we place several
kinematic cuts on the $z$, $\Phperp$ and $Q^2$ dependencies
in order
to isolate samples where the theoretical framework used in
this analysis is applicable.
To stay within the current fragmentation region, only data
for $z > 0.2$ are included, and to avoid contamination from
vector-meson production and soft-gluon effects, we exclude
data above $z = 0.6$.
For the $\Phperp$ dependence, we exclude the regions where $\Phperp$
is very small ($\Phperp > 0.2$~GeV) or very large ($\Phperp < 0.9$~GeV):
  the former to avoid acceptance issues for the lowest-$\Phperp$ bin
of the HERMES multiplicity data, and
  the latter to ensure the applicability of the Gaussian assumption,
without the need for introducing the $Y$ term~\cite{Collins:2016hqq}.
To stay above the charm threshold, we restrict ourselves to
$Q^2 > m_c^2$.

Because the existing SIDIS Collins asymmetry data have a rather small
$Q^2$ range, and $Q^2$ evolution effects tend to cancel in ratios,
there is no clear empirical indication of scale dependence in the
asymmetries.  It is a reasonable approximation, therefore, to neglect
the $Q^2$ dependence in the $F_{UT}^{\sin(\phi_h + \phi_s)}$ structure
function, and freeze the scale in the unpolarized $f_1^q$ and $D_1^q$
distributions in $F_{UU}$ at a value $Q^2 = 2$~GeV$^2$ that is typical
of SIDIS data.
(In contrast, since $e^+ e^-$ data are taken at higher energies,
neglecting the scale dependence between the $e^+ e^-$ and SIDIS
measurements would introduce uncontrolled errors from not including
the full TMD evolution where its effects may be important.)

In determining the transversity TMDs $h_1^q(x,k_\perp^2)$,
we parametrize the $x$ dependence by the form
	$h_1^q(x) = N_q x^{a_q} (1-x)^{b_q}$
for each of the flavors $q=u, d$ and $\bar q$,
assuming a symmetric sea,
	$h_1^{\bar{u}} = h_1^{\bar{d}} = h_1^s = h_1^{\bar{s}}$,
and use isospin symmetry to relate the distributions in
the proton and neutron.
For the Collins $\pi^\pm$ distributions, we use a similar
functional form to parametrize the $z$ dependence of the favored
	$H_{1\, \rm (fav)}^{\perp (1)} \equiv
	 H_{1\, \pi^+/u}^{\perp (1)} =
	 H_{1\, \pi^+/\bar d}^{\perp (1)}$
FFs and the unfavored
	$H_{1 \rm (unf)}^{\perp (1)}$
FFs for $\{ d, \bar u, s, \bar s \} \to \pi^+$, with the
distributions for $\pi^-$ related by charge conjugation.
For the $x$ dependence of the spin-averaged $f_1^q$ distributions
we use the CJ15 leading-order parametrization~\cite{Accardi:2016qay},
while for the $z$ depedence of $D_1^q$ we utilize the leading-order
DSS fit \cite{deFlorian:2007ekg}.
Choosing a different FF parametrization would not affect the results
significantly, as changes in the $z$ dependence of the FFs could be
compensated by modified widths in the Gaussian $\Phperp$ distributions.

For the transverse-momentum widths $\avkperp^q_f$ of the TMD PDFs
$f_1^q$ and $h_1^q$, two Gaussian widths are used, one for the
  valence type ($q=u, d$) and one for the
  sea-quark type ($q = \bar u, \bar d, s, \bar s$) functions.
Similarly, for the TMD FFs two Gaussian widths for $\avpperp^{h/q}_D$
are used, for the favored (such as $u$ or $\bar d$ to $\pi^+$) and
unfavored ($\bar u$ or $d$ to $\pi^+$) type of FF.
In total, we therefore have 23 parameters to be extracted
from data, 19 of which describe $F_{UT}^{\sin(\phi_h +\phi_s)}$
and 4 for the transverse part of $F_{UU}$.
To determine the latter, we perform an independent fit to the HERMES
$\pi^\pm$ and $K^\pm$ multiplicity data~\cite{Airapetian:2012ki},
which include 978 data points that survive the same cuts as employed
for $A_{UT}^{\sin(\phi_h+\phi_s)}$.

Using the nested sampling MC algorithm~\cite{Skilling:2004,
Mukherjee:2005wg, Shaw:2007jj}, we compute the expectation
value E[${\cal O}$] and variance V[${\cal O}$],
\begin{subequations}
\label{eq:EV}
\begin{eqnarray}
\hspace*{-0.5cm}
{\rm E}[{\cal O}]
&=& \int d^n a\, {\cal P}({\bm a}|{\rm data})\,
    {\cal O}({\bm a})\
\simeq\ \sum_k w_k\, {\cal O}({\bm a_k}),			\\
\hspace*{-0.5cm}
{\rm V}[{\cal O}]
&=& \int d^n a\, {\cal P}({\bm a}|{\rm data})
    \big( {\cal O}({\bm a}) - {\rm E[{\cal O}]} \big)^2		\notag\\
&\simeq& \sum_k w_k
	 \big( {\cal O}({\bm a_k}) - {\rm E[{\cal O}]} \big)^2,
\end{eqnarray}
\end{subequations}%
for each observable ${\cal O}$ (such as a TMD or a function of TMDs),
which is a function of the $n$-dimensional vector parameters ${\bm a}$
with probability density ${\cal P}({\bm a}|{\rm data})$
\cite{Sato:2016wqj}.
Using Bayes' theorem, the latter is given by
\begin{eqnarray}
{\cal P}({\bm a}|{\rm data})
&=& \frac{1}{Z}\, {\cal L}({\rm data}|{\bm a})\, \pi({\bm a}),
\end{eqnarray}
where $\pi({\bm a})$ is the prior distribution for the vector
parameters ${\bm a}$, and
\begin{eqnarray}
{\cal L}({\rm data}|{\bm a})
&=& \exp\left[ -\frac12 \chi^2({\bm a}) \right]
\end{eqnarray}
is the likelihood function, with
  $Z = \int d^n a\, {\cal L}({\rm data}|{\bm a})\, \pi({\bm a})$
the Bayesian evidence parameter.
Using a flat prior, the nested sampling algorithm constructs a set
of MC samples $\{\bm a_k\}$ with weights $\{w_k\}$, which are then
used to evaluate the integrals in Eqs.~(\ref{eq:EV}).

The results of the fit indicate good overall agreement with the
Collins $\pi^+$ and $\pi^-$ asymmetries, as illustrated in
\fref{dvt}, for both HERMES~\cite{Airapetian:2010ds} and
COMPASS ~\cite{Adolph:2014zba, Alekseev:2008aa} data,
with marginally better fits for the latter.
The $\chi^2/{\rm datum}$ values for the $\pi^+$ and $\pi^-$
data are 28.6/53 and 40.4/53, respectively, for a total of
$68.9/106 \approx 0.65$.  The larger $\chi^2$ for $\pi^-$
stems from the few outlier points in the $x$ and $z$ spectra,
as evident in \fref{dvt}.
The SIDIS-only fit is almost indistinguishable, with
$\chi^2_{\rm SIDIS} = 69.2$.
Clearly, our MC results do not indicate any tension between
the SIDIS data and lattice QCD calculations of $g_T$,
nor any ``transverse spin problem''.

\begin{figure}
\begin{center}
\hspace*{-0.2cm}\includegraphics[width=0.5\textwidth]{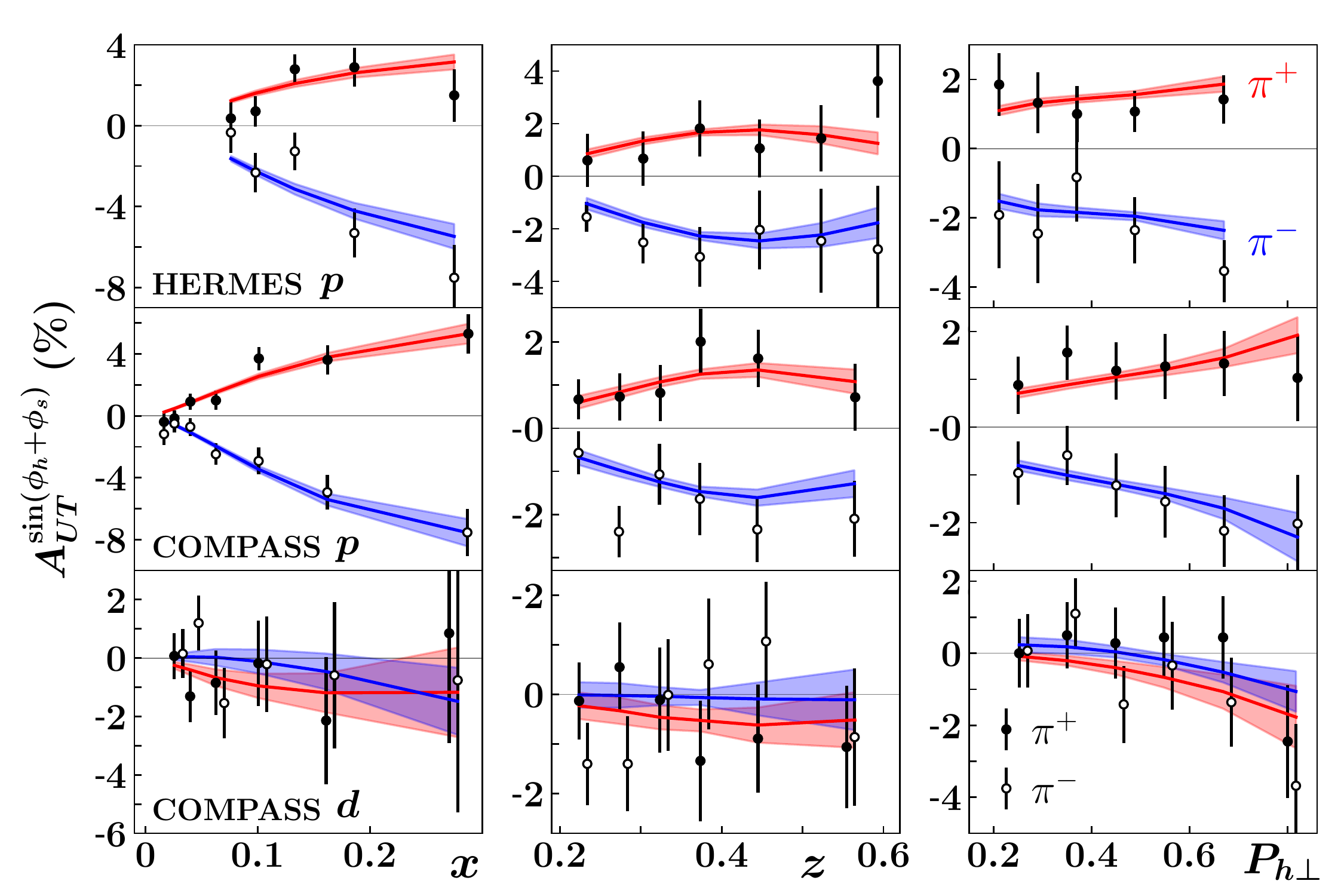}
\end{center}
\vspace*{-0.5cm}
\caption{Comparison of the full SIDIS$+$lattice fit with the
	$\pi^+$ (filled circles) and $\pi^-$ (open circles)
	Collins asymmetries $A_{UT}^{\sin(\phi_h+\phi_s)}$
	from HERMES~\cite{Airapetian:2010ds} and
	COMPASS~\cite{Adolph:2014zba, Alekseev:2008aa} (in percent),
	as a function of $x$, $z$ and $\Phperp$ (in GeV).}
\label{f.dvt}
\end{figure}

\begin{figure}
\begin{center}
\hspace*{-0.2cm}\includegraphics[width=0.5\textwidth]{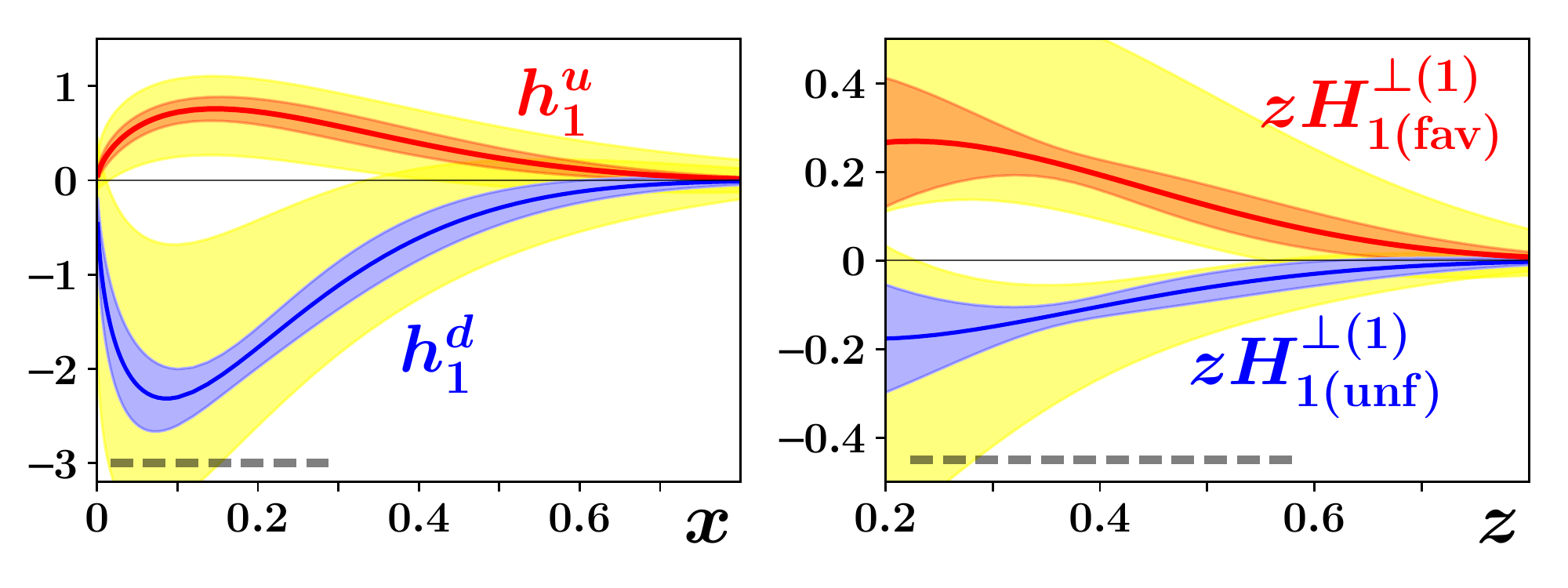}
\end{center}
\vspace*{-0.6cm}
\caption{Transversity PDFs $h_1^{u,d}$ and
	favored   $zH_{\rm 1 (fav)}^{\perp (1)}$ and
	unfavored $zH_{\rm 1 (unf)}^{\perp (1)}$ Collins FFs
	for the SIDIS$+$lattice fit (red and blue bands)
	at $Q^2=2$~GeV$^2$, compared with the SIDIS-only
	fit uncertainties (yellow bands).
	The range of direct experimental constraints is
	indicated by the horizontal dashed lines.}
\label{f.TC}
\end{figure}

The resulting transversity PDFs $h_1^u$ and $h_1^d$ and Collins
favored and unfavored FFs, $H_{\rm 1 (fav)}^{\perp (1)}$ and
$H_{\rm 1 (unf)}^{\perp (1)}$, are plotted in \fref{TC} for both
the SIDIS-only and SIDIS$+$lattice fits.
The positive (negative) sign for the $u$ ($d$) transversity PDF
is consistent with previous extractions, and correlates with
the same sign for the Collins FFs in the region of $z$ directly
constrained by data.
The larger $|h_1^d|$ compared with $|h_1^u|$ reflects the larger
magnitude of the (negative) $\pi^-$ asymmetry than the (positive)
$\pi^-$ asymmetry.
At lower $z$ values, outside the measured region, the uncertainties
on the Collins FFs become extremely large.
Interestingly, inclusion of the lattice $g_T$ datum has very
little effect on the central values of the distributions,
but reduces significantly the uncertainty bands.
The fitted antiquark transversity is consistent with zero,
within relatively large uncertainties, and is not shown in \fref{TC}.

For the transverse momentum widths, our analysis of the HERMES
multiplicities \cite{Airapetian:2012ki} gives a total
$\chi^2/{\rm datum}$ of 1079/978, with
  $\avkperp^q_{f_1} = 0.59(1)$~GeV$^2$ and 0.64(6)~GeV$^2$
for the unpolarized valence and sea quark PDF widths, and
  $\avpperp^{\pi/q}_{D_1} = 0.116(2)$~GeV$^2$ and 0.140(2)~GeV$^2$
for the unpolarized favored and unfavored FF widths.
These values are compatible with ones found in the analysis by
Anselmino~{\it et~al.}~\cite{Anselmino:2013lza} of HERMES and
COMPASS charged hadron multiplicities.
On the other hand, the similar values found for the sea
and valence PDF widths disagree with the chiral soliton
model~\cite{Schweitzer:2012hh}, for which the sea to valence
ratio is $\sim$~5.
Note also that while there appear some incompatibilities between the
$x$ dependence of the HERMES and COMPASS $P_{h\perp}$-integrated
$\pi^\pm$ multiplicities, our analysis uses only $P_{h\perp}$-dependent
HERMES data that are given in bins of $x$, $z$, $Q^2$ and $P_{h\perp}$.

The transverse momentum widths for the valence and sea transversity
PDFs are
  $\avkperp^q_{h_1} = 0.5(2)$~GeV$^2$ and 1.0(5)~GeV$^2$,
respectively, and
  $\avpperp^{\pi/q}_{H_1^\perp} = 0.12(4)$~GeV$^2$ and 0.06(3)~GeV$^2$
for the favored and unfavored Collins FF widths, respectively.
The relatively larger uncertainties on the $h_1$ and $H_1^\perp$
widths compared with the unpolarized widths reflect the higher
precision of the HERMES multiplicity data, and the order of magnitude
smaller number of data points for the Collins asymmetries.

\begin{figure}
\begin{center}
\hspace*{-0.2cm}\includegraphics[width=0.5\textwidth]{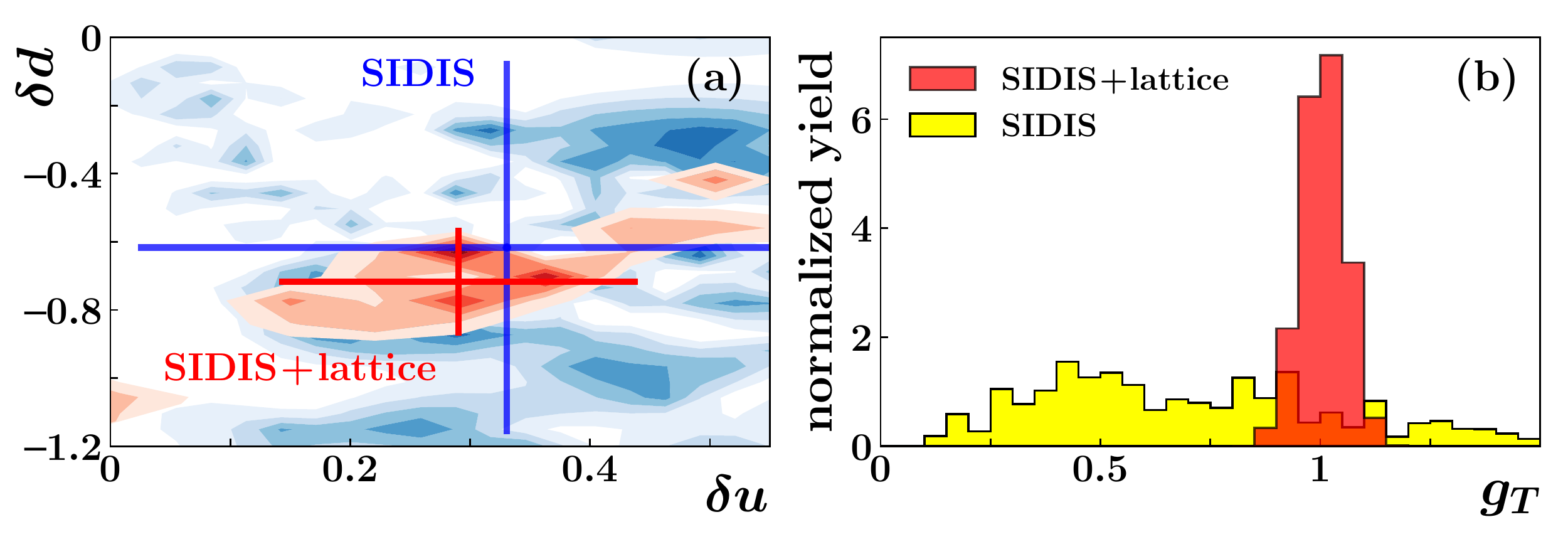}
\end{center}
\vspace*{-0.7cm}
\caption{(a)~Contour plot of $\delta u$ and $\delta d$ samples from
	the MC analysis, for the SIDIS only (blue) and
	SIDIS+lattice (red) analysis.  The expectation values and
	$1\sigma$ uncertainties for both fits are indicated by the
	respective error bars.
	(b)~Normalized yields for the isovector	tensor charge $g_T$,
	for the SIDIS-only (yellow histograms) and
	SIDIS+lattice (red histograms) MC analyses.}
\label{f.gT}
\end{figure}

Integrating the transversity PDFs over $x$, the resulting normalized
yields from our MC analysis for the $\delta u$ and $\delta d$ moments
are shown in \fref{gT}, together with the isovector combination $g_T$.
The most striking feature is the significantly narrower distributions
evident when the SIDIS data are supplemented by the lattice $g_T$ input.
The $u$ and $d$ tensor charges in \fref{gT}(a), for example, change from
	$\delta u =  0.3(3) \to  0.3(2)$ and
	$\delta d = -0.6(5) \to -0.7(2)$
at the scale $Q^2=2$~GeV$^2$, while the reduction in the uncertainty
is even more dramatic for the isovector charge in \fref{gT}(b),
	$g_T = 0.9(8) \to 1.0(1)$.
The earlier single-fit analysis of SIDIS data by
Kang~{\it et~al.}~\cite{Kang:2015msa} quotes
	$\delta u = 0.39(11)$ and
	$\delta d = -0.22(14)$, with
	$g_T = 0.61(25)$
at $Q^2 = 10$~GeV$^2$, in apparent tension with the lattice results.
This can be understood from \fref{gT}(b), which demonstrates that the
peak of the SIDIS-only distribution at $g_T \sim 0.5$ is consistent
with the lower values found in earlier maximum likelihood
analyses~\cite{Kang:2015msa, Ye:2016prn}, but does not give a good
representation of the mean value because of the long tail of the
$g_T$ distribution.

Future extensions of this work will explore incorporating TMD evolution
via the CSS framework \cite{Collins:1984kg, Collins:2011zzd}, and
improved treatment of the large-$\Phperp$ contributions through the
addition of the $Y$ term \cite{Collins:2016hqq}.
Inclusion of $K^\pm$ SIDIS and $e^+ e^-$ annihilation data will allow
further separation of sea quark flavor contributions to $h_1$ and
better constraints on the favored and unfavored Collins FFs.
Upcoming high-precision data from Jefferson Lab should also provide
significantly improved kinematical coverage at intermediate $x$ and
$z$ values.

%%%%%%%%%%%%%%%%%%%%%%%%%%%%%%%%%%%%%%%%%%%%%%%%%%%%%%%%%%%%%%%%%%%%%%%%
% Acknowledgments.
%
We are grateful to J.~Qiu for helpful comments.
This work was supported by the US Department of Energy contract
DE-AC05-06OR23177, under which Jefferson Science
Associates, LLC operates Jefferson Lab, and by the National Science
Foundation contracts PHY-1623454, PHY-1653405 and PHY-1659177.

%%%%%%%%%%%%%%%%%%%%%%%%%%%%%%%%%%%%%%%%%%%%%%%%%%%%%%%%%%%%%%%%%%%%%%%%
\bibliography{paper.bbl}

\end{document}